\documentclass[pre,floatfix,twocolumn,showpacs,superscriptaddress]{revtex4-1}

\usepackage{graphicx}
\usepackage{bm}

\newcommand{\ve}[1][K]{\mathbf{#1}}

\usepackage{amsmath}
\usepackage{amssymb}
\usepackage{latexsym}
\usepackage{multirow}
\usepackage{xcolor}
\usepackage{hyperref}

\begin{document}

\title{Universal kinetics of imperfect reactions in confinement}

\author{Thomas Gu\'erin}
\affiliation{Laboratoire Ondes et Mati\`ere d'Aquitaine, CNRS/University of Bordeaux, F-33400 Talence, France}

\author{Maxim Dolgushev}
\affiliation{Laboratoire de Physique Th\'eorique de la Mati\`ere Condens\'ee, CNRS/Sorbonne University, 
 4 Place Jussieu, 75005 Paris, France}
 
\author{Olivier B\'enichou}
\email{benichou@lptmc.jussieu.fr}
\affiliation{Laboratoire de Physique Th\'eorique de la Mati\`ere Condens\'ee, CNRS/Sorbonne University, 
 4 Place Jussieu, 75005 Paris, France}

\author{Rapha\"el Voituriez}
\affiliation{Laboratoire de Physique Th\'eorique de la Mati\`ere Condens\'ee, CNRS/Sorbonne University, 
 4 Place Jussieu, 75005 Paris, France}
\affiliation{Laboratoire Jean Perrin, CNRS/Sorbonne University, 
 4 Place Jussieu, 75005 Paris, France }

\bibliographystyle{naturemag}
 \renewcommand{\refname}{References}
 \def\bibsection{\section*{\refname}} 

\date{\today}

\begin{abstract} 
Chemical reactions generically require that particles come into contact. In practice, reaction is often imperfect and can necessitate multiple random encounters between reactants. In confined geometries, despite notable recent advances, there is to date no general analytical treatment of such imperfect transport–limited reaction kinetics. Here, we determine the kinetics of imperfect reactions in confining domains for any diffusive or anomalously diffusive Markovian transport process, and for different models of imperfect reactivity. We show that the full distribution of reaction times is obtained in the large confining volume limit from the knowledge of the mean reaction time only, which we determine explicitly. This distribution for imperfect reactions is found to be identical to that of perfect reactions upon an appropriate rescaling of parameters, which highlights the robustness of our results. Strikingly, this holds true even in the regime of low reactivity where the mean reaction time is independent of the transport process, and can lead to large fluctuations of the reaction time – even in simple reaction schemes. We illustrate our results for normal diffusion in domains of generic shape, and for anomalous diffusion in complex environments, where our predictions are confirmed by numerical simulations. \end{abstract}

\maketitle

\section*{Introduction}  
  
The First Passage Time (FPT) quantifies the time needed for a random walker to reach a target site~\cite{Redner:2001a,Condamin2007,pal2017first,grebenkov2016universal,benichou2010optimal,vaccario2015first,metzler2014first,Schuss2007,newby2016first,ReviewBray}. This observable is involved  in various areas of biological and soft matter physics and is particularly relevant in the context of reaction kinetics, because two reactants  have to meet before any reaction can occur~\cite{RiceBook,Berg1985,lindenberg2019chemical}. When the reaction is \textit{perfect}, i.e. when it occurs for certain upon the first encounter, its kinetics is controlled by the first passage statistics of one reactant, described as a random walker, to a target site. Of note, earlier works have determined the mean~\cite{Condamin2007,Condamin2005,Schuss2007}  and the full asymptotic distribution~\cite{Benichou2010,godec2016universal}  of first-passage times in confinement for broad classes of transport processes.

While most of the literature focuses on perfect reactions, the case of \textit{imperfect} reactions (i.e. which do not occur with certainty upon the first encounter) arises in a variety  of contexts (see \cite{grebenkov2019imperfect} for a recent review) :  if  reaction occurs only when  reactants meet with  prescribed  orientations~\cite{Berg1985} or after crossing an energy~\cite{shoup1982role} (or entropy~\cite{zhou1991rate}) activation barrier, if the target site is only partially covered by reactive patches \cite{berg1977physics}, or in the case of gated reactions where  the target (or the reactant) switches between reactive and inactive states~\cite{reingruber2009gated,benichou2000kinetics}. 

The formalism to calculate the rate of imperfect reactions between  diffusive spherical particles in the dilute regime (thus in infinite space) is well established \cite{collins1949diffusion,doi1975theory,Berg1985,traytak2007exact}. However, geometric confinement has proved to play an important role in various contexts, such as reactions in microfabricated reactors or  in cellular compartments. 
Yet, the kinetics of imperfect reactions in a confined volume  is still only partially understood:  existing methods are  restricted to (i) diffusive (or amenable to diffusive) transport  processes \cite{isaacson2016uniform,isaacson2013uniform,lindsay2017first,mercado2019first}
 and most of the time   (ii) specific shapes of confining volume \cite{grebenkov2010searching,grebenkov2017effects,grebenkov2018strong,grebenkov2018towards} (spherical or cylindrical). In fact, a general theoretical framework to quantify the kinetics of imperfect reactions involving  non Brownian transport (such as anomalous diffusion in complex  environments \cite{kopelman1988fractal}) in general confined domains is still missing.  
 
 \begin{figure}[h!]
\includegraphics[width=9cm]{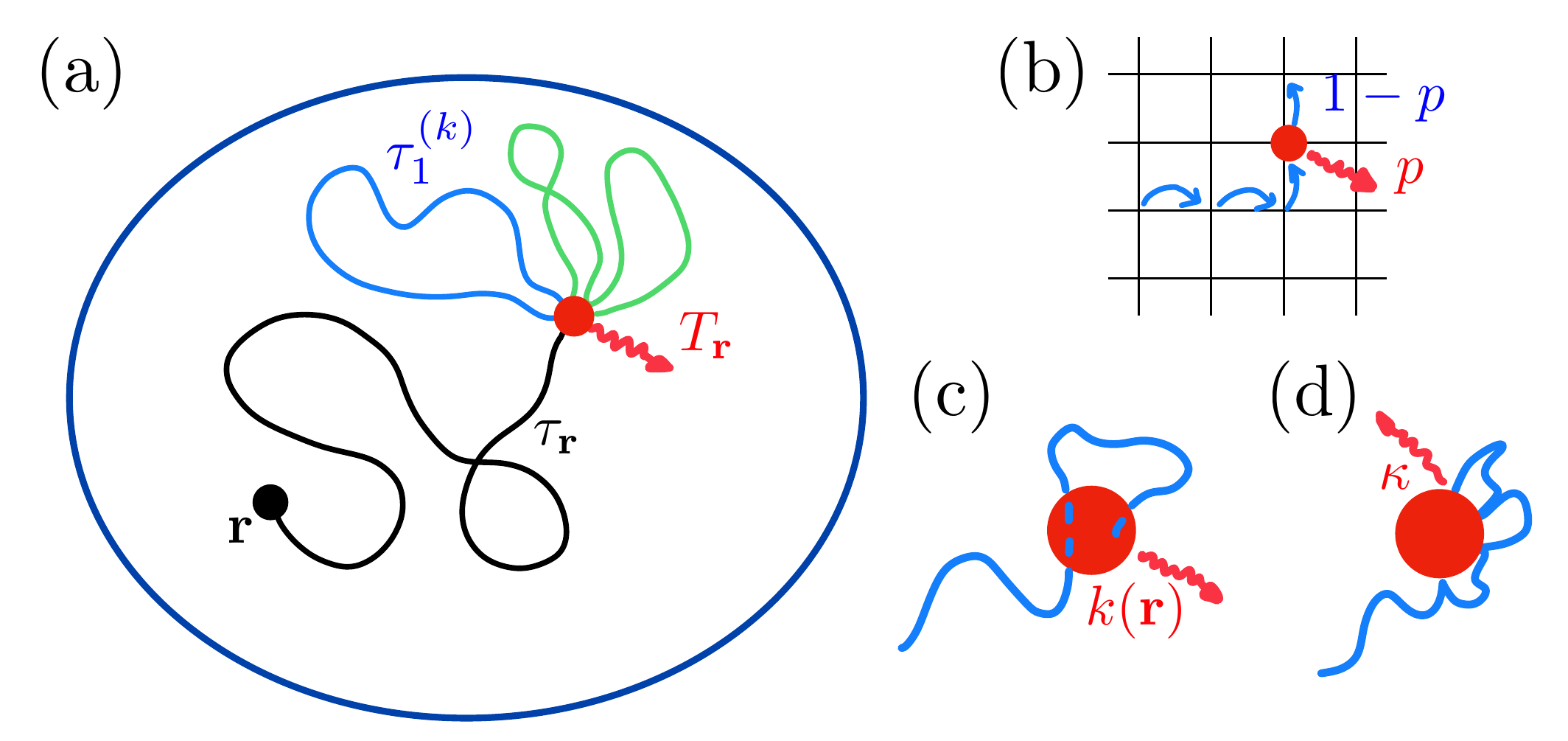} 
\caption{ {\textbf{ Imperfect reaction kinetics in confinement. }} (a) In the case of imperfect reactions, multiple random interaction events between reactants are typically required before reaction occurs. The reaction time $T_{\bf r}$ for a random walker starting from ${\bf r}$  with a target (red dot) can then be written $T_{\bf r}= \tau_{\bf r} + \sum_{k=2}^{n}\tau^{(k)}_1$, where $\tau_{\bf r}$ is the first-passage time (FPT) to the target, $n$ is the total number of visits to the target before reaction, and $\tau^{(k)}_1$ is a  first return time to the target. (b) In the case of discrete space models, imperfect reactivity is parametrized by the probability $p$ that reaction occurs at each visit of the random walker to  the target. In the case of continuous space models, imperfect reactivity is modeled either by (c) a reaction rate $k({\bf r})$ when the random walker is within the reactive volume that defines the target, or (d) partially absorbing boundary conditions (parametrized by  $\kappa$) at the target boundary.  }
\label{Figdessin}
\end{figure}

Here, we propose a  formalism that  determines the full kinetics of imperfect reactions in confinement  for general Markovian processes in the large confining volume limit (see Fig. \ref{Figdessin}). This allows us  to answer the following questions: (i) Is reaction limited by transport or reactivity ? (ii) What is the magnitude of the fluctuations of the reaction time ? In particular, is the first moment sufficient to fully determine reaction kinetics  ? (iii) Do reaction kinetics depend on the choice of model  of imperfect reactivity-- namely  partially reflecting (Robin) conditions \cite{collins1949diffusion,Szabo1980,sano1979partially} or sink with locally uniform absorption rate \cite{doi1975theory,WILEMSKI1974a} in continuous models, or finite  reaction probability in discrete models ?

\section*{Results and Discussion}
  
\textit{Discrete model of imperfect Reactions.-} A first straightforward definition of imperfect reactivity is based on the statistics of encounter events between reactants, and thus  requires a discrete description of the dynamics. We therefore start  by considering  a Markovian random walker moving on a discrete space (or network)  of $N$ sites. We consider a continuous time dynamics with exponentially distributed waiting times on each site, where  $\nu_i$ denotes the jump rate from site $i$ to any neighboring site. The  reactive site is denoted $i=0$. Imperfect reactivity is then naturally defined as follows : each time the walker visits the reactive site, reaction occurs with probability $p$, and the random walk continues without reaction with probability $1-p$.  We call $T_{\ve[r]}(p)$ the \textit{reaction time} and $F(T\vert \ve[r],p )$ its probability density function (PDF) for a random walker starting from   ${\bf r} $. Next, we call $\tau_{\ve[r]}$ the \textit{first passage time} to the reactive site starting from ${\bf r} $ (including the residence time on the reactive site), and we call $F^*(\tau_{\ve[r]} \vert \ve[r])$ its PDF. We also introduce the \textit{first return time} to the target $\tau_1$ (i.e. the first passage time to the target starting from any site at distance 1 from the target) and $F_1^*(\tau_1)$ its PDF.  The probability that a reaction happens after exactly $n$ visits to the target is given by  $p(1-p)^{n-1}$, in which case $T_{\ve[r]}(p)$ is the sum of the first passage time (starting from $\ve[r]$) and of $n-1$ independently distributed first return times (see Fig \ref{Figdessin}). Hence, partitioning over the number of visits $n$ yields 
\begin{align}\label{F_de_t}
 F(T  & \vert \ve[r]  ,p )=
  \sum_{n=1}^{\infty}  \int_0^\infty\mathrm{d}\tau_{\ve[r]} \left[\prod_{k=2}^n \int_0^\infty d\tau_1^{(k)} F_1^*(\tau_1^{(k)})  \right]
 \nonumber\\
 \times & p(1-p)^{n-1}  F^*(\tau_{\ve[r]}  | \ve[r]) \delta\left(T-\tau_{\ve[r]}-\sum_{k=2}^n   \tau_1^{(k)}  \right),
 \end{align}
 where $\tau_1^{(k)}$ represents the return time after $k-1$ visits to the reactive site.  This exact equation is conveniently rewritten after Laplace transform  (denoted  $\tilde{f}(s)=\int_0^\infty dt  f(t)e^{-st} $ for any function $f$):
 \begin{equation}
\tilde{F}(s\vert \ve[r],p)=\frac{p \ \tilde{F}^*(s\vert \ve[r])}{1-(1-p)\tilde{F}_1^*(s)   }. \label{ClosedFormFPTDensity}
\end{equation}
(see Supplementary Note 1 for details). In the small $s$ limit, the property $\tilde{F}(s\vert \ve[r],p)\simeq 1-s \langle T_{\ve[r]}(p)\rangle$ can be used to obtain an exact expression  of the mean reaction time as a function of the mean first passage and the mean return time:
\begin{equation}
\langle T_{\ve[r]}(p)\rangle=\langle\tau_{\ve[r]} \rangle+\frac{1-p}{p} \langle \tau_1\rangle   \label{meanRT} .
\end{equation}
Of note,  expression \eqref{meanRT} [as well as \eqref{ClosedFormFPTDensity}]  is a straightforward consequence of well known results on random sums \cite{feller}, bearing here a clear interpretation because $(1-p)/p$ is the mean number of encounter events. Below, we make this result fully explicit by determining $\langle \tau_{\ve[r]}\rangle$ and $\langle \tau_1\rangle$. 

The mean return time $\langle \tau_1\rangle$ can be obtained exactly from the knowledge of the stationary probability density $q_i$ for the random walker to be at site $i$ in absence of target ; this exact result is known as Kac theorem \cite{aldousFill2014} and yields 
\begin{equation}
\langle \tau_1\rangle = \frac{1}{q_0\nu_0}=\frac{N}{\nu_0},\label{Kac}
\end{equation}
where we have chosen a uniform stationary distribution $q_i=1/N$, which is realized when the waiting time $1/\nu_i$ at each site  is inversely proportional to the number of neighbors \cite{masuda2017random}.

To gain explicit insight of the behavior of the first reaction times, we next determine $\langle \tau_{\ve[r]}\rangle$ and make use of the scale invariance property observed for a broad class of random walks, for which one can define a fractal space dimension $d_f$ (defined such that the characteristic size $R$ grows as $R\propto N^{1/d_f}$) and a walk dimension $d_ w$ such that the mean square displacement of a random walker scales as $\langle r^2(t)\rangle\propto t^{2/d_w}$ (without absorption, in unconfined space). Here, we make use of the chemical distance $r$, defined as  the minimal number of links between two sites.  The first passage kinetics is known to strongly differ for compact walks ($d_w>d_f$, for which the random walker explores densely its surrounding space and the probability to visit a site in infinite space is one) or non-compact walks $(d_w<d_f$, for which an infinite trajectory typically leaves a fraction of unvisited sites which is almost surely one).
We shall prove here that the effect of imperfect reactivity is markedly different in these two cases as well.
 
\begin{figure}[h!]
\includegraphics[width=9cm]{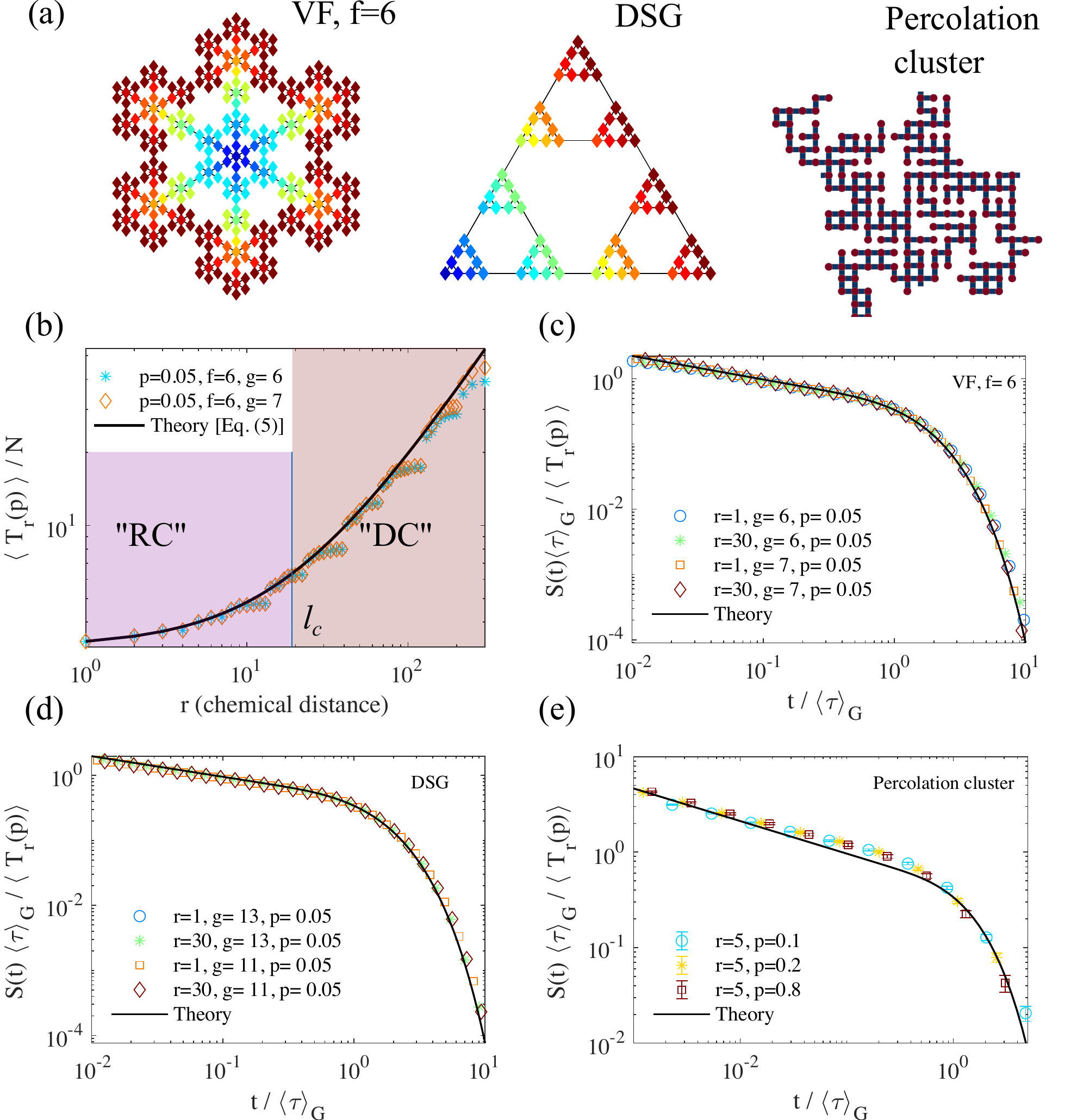} 
\caption{{\textbf{ Distributions of  reaction times for compact transport. } (a) Fractal networks for which we performed simulations:  Vicsek fractal (VF), here of functionality $f=6$, dual Sierpinski gasket (DSG), and two-dimensional percolation cluster (a 2D network in which half the bonds are randomly suppressed). For the VF and DSG, the color codes   the distance to the reactive site, taken at an apex for DSG and at the center for VFs. (b) Mean Reaction time for VF and an absorption probability $p=0.05$. Reaction Controlled (RC) and Diffusion controlled (DC) regimes appear respectively below and above the length $l_c$ defined in Eq.~(\ref{lcDef}). (c),(d),(e): Survival probabilities, in rescaled coordinates for various generations $g$ and initial distances $r$ for (c): VF, $f=6$,  (d): DSG and (e): percolation cluster extracted from a $200\times200$ two dimensional square lattice. Details of numerical procedures and Additional examples of fractals can be found in Supplementary Note 3, Supplementary Figures S1-S2, and Supplementary Table S1. In (d) the error-bars represent $95\%$ confidence intervals.}}
\label{FigFractals}
\end{figure}

We first focus on the compact case $d_w>d_f$, for which it  was shown~\cite{benichou2008zero} that  $\langle\tau_r\rangle \simeq \alpha N r^{d_w-d_f}$ for large $r$ and large volume $N$, where $\alpha$ is a constant independent of $N$ and $r$. Following \cite{benichou2008zero}, we assume that this scaling relation holds up even for $r=1$. Making use of  the above determination of $\langle \tau_1\rangle$, this  yields $\alpha =1/\nu_0$.   This leads to the following fully explicit determination of the mean reaction time:
 \begin{equation}
\langle T_r(p) \rangle \simeq \frac{N \ r^{d_w-d_f}}{\nu_0}+\frac{N (1-p)}{p\ \nu_0} .
\end{equation}
As expected, the reaction time is thus the sum of a diffusion controlled (DC) time $\langle \tau_r \rangle$, obtained when $p=1$, corresponding to the time needed for the reactants to meet, and a reaction controlled (RC) time $\langle \tau_1\rangle (1-p)/p$, which dominates when $p\to 0$, corresponding to the sequence of returns to the target  needed for reaction to occur.   These two times are equal when $r\simeq l_c$, where the characteristic distance $l_c$ is given by 
\begin{equation}
l_c=[(1-p)/p]^{1/(d_w-d_f)}. \label{lcDef}
\end{equation}
For this compact case, we can therefore split the confining domain into a region where the reaction time is reaction controlled (RC, for $r<l_c$), and another one where it is diffusion controlled (DC, $r>l_c$), see Fig. \ref{FigFractals}(b).  Remarkably, we note that DC region disappears only when the size $R$  of the confining volume  becomes of the order of $l_c$, \textit{i.e.} when $p\ll 1/R^{d_w-d_f}$; this means that  even for very small values of the intrinsic reactivity there will exist DC regions for large enough volumes. 

To quantify reaction kinetics at all time scales, the full distribution of the reaction time, or equivalently  the survival probability $S(t \vert \ve[r],p)$, defined as the fraction of walkers that have not reacted up to time $t$, is needed. We show in Supplementary Note 2 how to determine $S(t)$ by evaluating the leading order behavior of all the moments $\langle T_r^n(p)\rangle$ in the large volume limit (defined by $N\to \infty$ with all other parameters fixed). Using the additional hypothesis that the scaling behavior of of all moments $\langle\tau_r^n\rangle\sim r^{d_w-d_f}R^{d_f+(n-1)d_w}$ holds up to $r=1$, this leads to  an explicit determination of the survival probability 
\begin{equation}\label{S_scaling}
S(t\vert \ve[r],p) \simeq\frac{\langle T_r(p) \rangle}{ \langle \tau \rangle_{\text{\tiny G}} }\ \Phi_\nu\left(\frac{t}{\langle \tau \rangle_{\text{\tiny G}}}\right),
\end{equation}
where $\langle \tau \rangle_{\text{\tiny G}} $ is the global mean first passage time, \textit{i.e.} the average of $\langle \tau_{\ve[r]} \rangle $ over all starting positions of the random walker and is independent of $p$. Here, $\Phi_{\nu}$ is a universal function  depending only on $\nu=d_f/d_w$, which was obtained \cite{Benichou2010} for the first passage problem by relying in the O'Shaughnessy-Procaccia operator \cite{OShaughnessy1985} (which is known to provide accurate expressions for propagators for not-too-large distances \cite{klafter1991propagator}):
\begin{equation}\label{s_theta}
\Phi_\nu(\theta)= \sum_{k=0}^{\infty}\frac{J_{\nu}(\alpha_k)\alpha_k^{1-2\nu}\Gamma(\nu)2^{2\nu}\nu^2}{J_{1-\nu}(\alpha_k)\Gamma(2-\nu)(1+\nu)}e^{ -\frac{\alpha_k^2\nu}{2(1-\nu^2)}\theta}.
\end{equation}
Here $\Gamma$ is the gamma function, $J$ is the Bessel function of the first kind, and $\alpha_0<\alpha_1<...$ are the zeros of the function $J_{-\nu}$. 
 
 Several comments are in order. (i) This  main result shows that  the functional form of the survival probability is exactly the same as that of the first passage time to the target (obtained for $p=1$), with a rescaled prefactor $\langle T_r (p)\rangle$ that encompasses all the dependence on the reactivity parameter $p$. It generalizes the result obtained for perfect reactions \cite{Benichou2010}. (ii) Importantly, the full  distribution can be obtained from the knowledge of the first moment $\langle T_r (p)\rangle$ \textit{only}, which makes the mean the key quantity to determine reaction kinetics. (iii) Remarkably, the shape of the reaction time distribution is the same as that of first passage times even in regions of the domain where the mean reaction time is  reaction controlled and seemingly independent of the dynamics. As stressed above, the dependence on $p$ lies only in the prefactor of the survival probability. This implies that the property of broadly distributed reaction times (non-exponential), characteristic of first passage times for compact transport, is maintained \textit{even for low intrinsic reactivity}  in large networks. (iv) The importance of fluctuations can be quantified by  the ratio $\langle T^2\rangle/\langle T\rangle^2\sim R^{d_w}/\langle T\rangle\gg1$, which is large in the large volume limit that we consider. 

In order to test these predictions for compact processes, we have performed numerical calculations on different examples of both disordered and deterministic fractal networks: the 2-dimensional critical percolation cluster,  the Vicsek fractals and the  dual Sierpinski gasket, see Fig.\ref{FigFractals}(a). This enables us to test different values of $d_f,d_w$. This class of  models has been  used to describe transport in disordered media —for example in the case of  anomalous diffusion in crowded environments like biological cells \cite{malchus2010elucidating,saxton2008biological,Benichou:2011} —as a first step to account for geometrical obstruction and binding effects
involved in real crowded environments. Our calculations of the reaction times in the case of deterministic fractals are based on a recursive construction of the eigenvalues and eigenfunctions of the connectivity matrix (see Supplementary Note 3 for details) and enable us to obtain exact forms for the Laplace transform of $S(t)$ for volumes up to $N\sim 10^6$ sites. As seen on Fig.\ref{FigFractals}, these numerical results confirm our predictions for the evaluation of the mean first passage time and the rescaled form of the survival probability. These results indicate that our approximations (i.e. the use of the O'Shaughnessy-Procaccia operator, the hypothesis that scaling of all moments hold up to $r=1$, large volume limit) lead to  accurate predictions for the mean reaction time and its full distribution. Of note, even for small values of the  reaction probability ($p=0.05$) the shape of reaction time distribution is  exactly the same as that of first passage times,  as we predict. In the limit of small $p$ (at fixed volume), one expects that the reaction becomes much slower than the transport step, with an exponentially distributed reaction time. However,  this exponential regime appears when the length $l_c$ in Eq. (\ref{lcDef}) becomes comparable to the size $R$ of the fractal, i.e. when $p\ll p^*\equiv 1/N^{d_w/d_f-1}$. Since $p^*$ vanishes for large $N$, this means that the distribution of first passage times remains broadly distributed, with no well defined reaction rate  even for very low values of   $p$. 

\textit{Non-compact case.-} We now focus on  non-compact processes ($d_w<d_f$).  In this case, we make use of   the asymptotic FPT distribution, which can be written \cite{Benichou2010} as  
\begin{equation}
F^*(t\vert r)=\left(1-\frac{\langle \tau_r\rangle}{\langle \tau \rangle_{\text{\tiny G}}}\right) \delta(t)+\frac{\langle \tau_r\rangle}{\langle \tau \rangle_{\text{\tiny G}}^2} e^{-t/ \langle\tau \rangle_{\text{\tiny G}} }, \label{FPTDensityNonCompact}
\end{equation} 
where $ \langle \tau \rangle_{\text{\tiny G}}$ has been defined above.  The term $\delta(t)$ accounts for the FPT density restricted to trajectories that do not reach the boundary before finding the target, the shape of the function approximated by this $\delta$-function does not modify the value of the moments of the distribution in the large volume limit. 
Now, we make use of this  separation of time scales in the FPT distribution and obtain finally  the distribution of the reaction time  by (i) taking the Laplace transform  of (\ref{FPTDensityNonCompact}), (ii) inserting the result  into (\ref{ClosedFormFPTDensity}) and (iii) taking the inverse Laplace transform. The result of this procedure for the survival probability is 
\begin{equation}
S(t\vert r,p) = \frac{\langle T_r(p) \rangle}{ \langle T(p) \rangle_{\text{\tiny G}}} e^{-t /  \langle T(p) \rangle_{\text{\tiny G}} },\label{SurvNonCompact}
\end{equation}  
where the mean reaction time $\langle T_r(p) \rangle$ is deduced from Eqs. (\ref{meanRT}),(\ref{Kac}), and $ \langle T(p) \rangle_{\text{\tiny G}} = \langle T_{r=\infty} \rangle$ is the global (indexed by G) mean reaction time, i-e averaged over all starting positions.   
This result has important consequences.  (a) Similarly to the compact case, the shape of the survival probability for imperfect reactions is the same as that of first passage times, with renormalized parameters; in particular  the mean gives access to the full distribution, and is thus the key quantity to quantify reaction kinetics, as in the compact case. (b) Because  the mean FPT scales as $\langle \tau\rangle\sim (N/\nu_0 )g(r)$, where $g$ is a bounded function of $r$,   the mean reaction time is dominated by the reaction limited step in the full domain as soon as $p\ll 1$ for any domain size, in contrast to the compact case.   
(c) Note that, in Eq. (\ref{SurvNonCompact}) one has $S(t\to0)<1$, which means that it does not take into account the events whose duration does not scale with $R$ ; the survival probability for these events was identified to the survival probability in infinite space \cite{Benichou2010,grebenkov2018strong}. Nevertheless, Eq. (\ref{SurvNonCompact}) can be used to calculate \textit{all the  moments} $\langle T_r^n(p)\rangle $ with $n\ge1$ in the large volume limit. 

\textit{Continuous models : imperfect extended targets.- } We  now aim at discussing alternative microscopic models of imperfect reactivity, which are naturally defined in continuous space.  We consider a $d$-dimensional Brownian diffusive particle of diffusion coefficient $D$ and analyze two classical models of imperfect targets (see Fig. \ref{Figdessin}) : (i) a sink region $V_r$ (in which the reaction happens with rate $k(\ve[r])$ and vanishes elsewhere) and (ii) a target region $S_r$ with partially reactive impenetrable boundary. Our above results for discrete models show that the full distribution of reaction times can be obtained in the large volume limit from the first moment of the reaction time  only ; we conjecture and verify numerically  that this holds also for continuous models. We are thus back to determining the mean reaction time  in both cases (i) and (ii).  In case (i),  the mean reaction time $\langle T(\ve[r])\rangle$ starting from the position $\ve[r]$  satisfies the following backward equation \cite{gardiner1983handbook}:  
\begin{equation}\label{BE}
[D\Delta_{\ve[r]}-k(\ve[r])]\langle T(\ve[r])\rangle=-1.
\end{equation}
We next define $\Phi(\ve[r])=\lim_{V\to\infty} \langle T(\ve[r])\rangle/V$, and obtain from \eqref{BE} (and  \eqref{BE}  integrated over the volume) :
\begin{equation}
[D\Delta_{\ve[r]}-k(\ve[r])]\Phi(\ve[r])=0, \ \int d\ve[r]\  k(\ve[r])\ \Phi(\ve[r])=1,\label{MFPTSink}
\end{equation}
which fully determines $\Phi$ for all $\ve[r]\in{\mathbb R}^d$. This formalism can be adapted to the case (ii) of partially reactive target of surface $S_r$ characterized by a  surface reactivity $\kappa$ that interpolates from perfect reaction ($\kappa\to \infty$) to complete absence of reaction ($\kappa\to 0$) \cite{collins1949diffusion,Szabo1980,sano1979partially,grebenkov2019imperfect}; we obtain  
\begin{equation}
\Delta_{\ve[r]} \Phi\vert_{\ve[r]\in{\mathbb R}^d\backslash S_r}=0,\    D \partial_n\Phi =\kappa\Phi |_{\ve[r]\in S_r},   \   \int_{S_r} dS  \kappa \Phi =1.\label{MFPTRobin}
\end{equation}
 These equations (\ref{MFPTSink}) and (\ref{MFPTRobin}) generalize the formalism of Ref. \cite{Benichou2008} to the case of imperfect reactions. Note that (i) they can be extended to general Markovian transport operators,  for both compact and non-compact cases, and (ii) they are valid   for any shape of the confining volume. 
 
\begin{figure}[h!]
 \includegraphics[width=9cm]{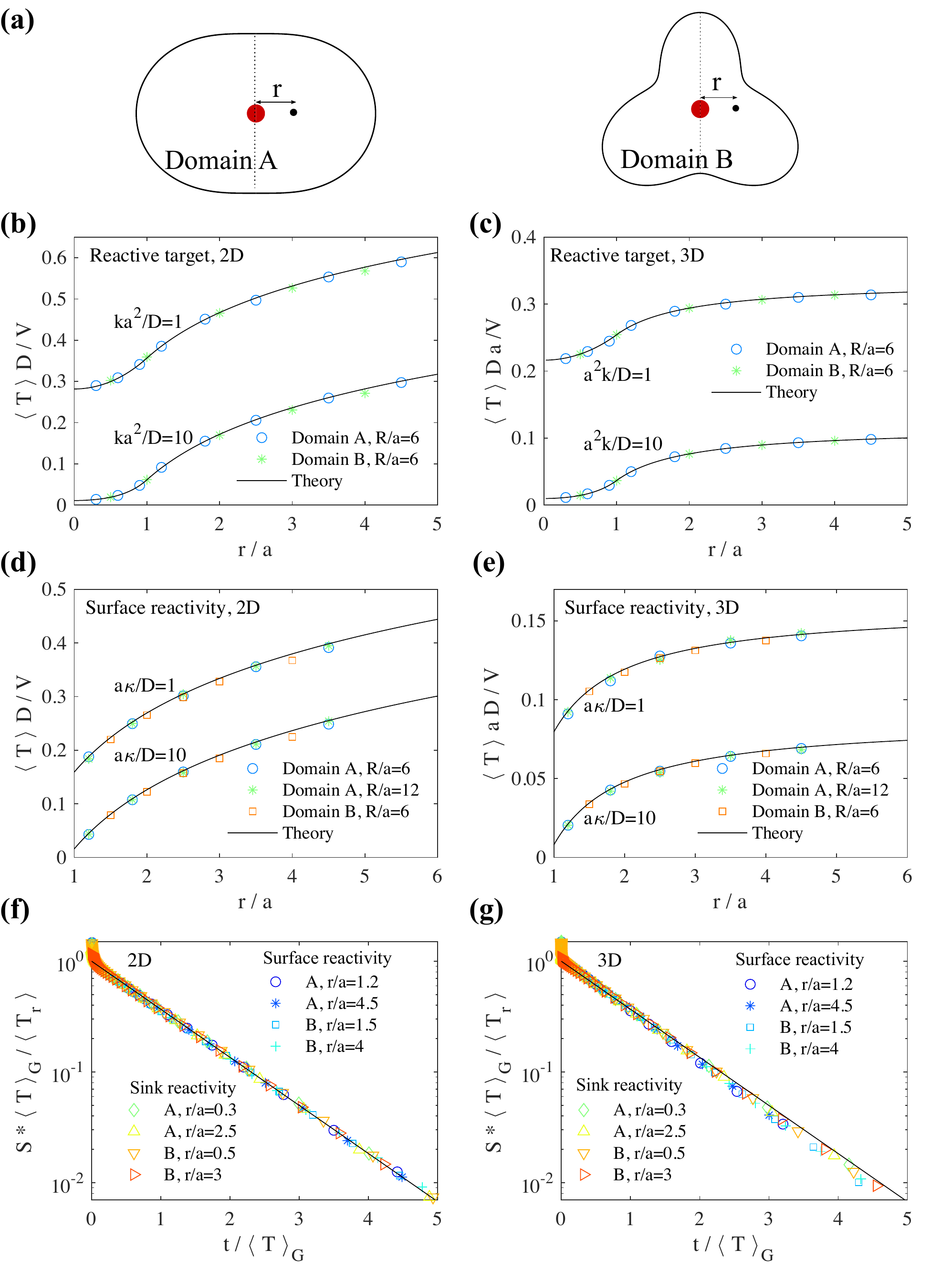} 
\caption{{\textbf{Distribution of reaction times for diffusive processes in various confining domains, for sink and surface reactivity.}} (a) Geometry of the confining domains (called $A$ and $B$) that are considered for stochastic simulations. In 2D, these domains are defined in polar coordinates by $r(\theta)=R f(\theta)$ with $f=1.6(1+0.5\cos^2\theta)$ for domain $A$ and $f=1.6(1+0.1\sin\theta +0.3\sin3\theta)$ for domain $B$. Domains in 3D are obtained by considering revolution of 2D surfaces around the vertical dashed line. The geometry of the target (red sphere) and initial position of the random walker are indicated. In the figure, we have used $R=6a$. (b),(c),(d),(e) Results of stochastic simulations for the mean reaction time in 2D/3D, for surface/sink reactivity, compared to our theoretical expressions. (f) and (g) Rescaled survival probabilities for 2D/3D simulations, all parameters are in legend except for $R/a=6$ and $ka^2/D=1$ (for sink reactivity) and $\kappa a/D=1$ (for surface reactivity). In 3D we evaluated $\langle T\rangle_{\text{\tiny G}}=V\phi(\infty)$. In 2D, we used $\langle T\rangle_{\text{\tiny G}}=V\phi(1)+\langle\tau\rangle_{\text{\tiny G}}$ where $\langle\tau\rangle_{\text{\tiny G}}$ was evaluated numerically for each domain. In all simulations we used a time step $\Delta t=10^{-4}a^2/D$. For surface reactivity we implemented our simulation algorithm by using Ref. \cite{singer2008partially}. Error-bars ($95\%$
 confidence intervals) are smaller than symbols. }
\label{FigDiffusion}
\end{figure} 
 
To illustrate our formalism, we give solutions for diffusive transport for dimensions $d=1,2,3$ for a spherical target of radius $a$. For the case (i) of a  sink region $k(r)=k\theta(a-r)$ with $\theta$ the Heaviside step function, the MRT outside the sink region ($r>a$) reads 
\begin{equation}
\frac{\langle T\rangle  }{V}= \frac{1}{D}
\begin{cases}
\frac{r}{2}+\frac{a}{2}\left[\frac{\cosh(\sqrt{K})}{\sqrt{K} \sinh(\sqrt{K})}-1\right] & (d=1) \\
\frac{1}{2\pi}\ln\frac{r}{a}+\frac{I_0\left(\sqrt{K}\right)}{2\pi \sqrt{K} I_1\left(\sqrt{K}\right)} & (d=2)\\
-\frac{1}{4\pi r }+\frac{\sqrt{K} }{4\pi a [\sqrt{K} -\tanh(\sqrt{K})]} & (d=3)
\end{cases}\label{TheorSink}
\end{equation}
where $K=ka^2/D$ and $I_0,I_1$ are modified Bessel functions of the first kind. In the case (ii) of a  partially reactive impenetrable spherical target we obtain
\begin{equation}
\frac{\langle T\rangle}{V}=
\begin{cases}
\frac{1}{2D} (r-a) + \frac{1}{2\kappa} & (d=1)\\
\frac{1}{2\pi D}\ln(r/a)+\frac{1}{2\pi a \kappa} & ( d=2)\\
\frac{1}{4\pi Da}-\frac{1}{ 4\pi D r}+\frac{1}{ 4\pi a^2\kappa} & (d=3)\label{TheorRobin}
\end{cases}
\end{equation} 
Of note, in $d=3$, the MRT at $r=\infty$ in Eqs. (\ref{TheorRobin},\ref{TheorSink}) is the inverse of effective reactions rates calculated in \cite{collins1949diffusion,doi1975theory}, and in fact the expression (\ref{SurvNonCompact}) then corresponds to the survival probability at long times for $r/a\gg1$ identified in Ref. \cite{isaacson2016uniform}.   
Finally, our results show that both models are equivalent in the low reactivity limit upon the identification $\kappa S_r=k V_r$; however, in the high reactivity limit,  the RC  time scales as $\kappa^{-1}$ for surface reactivity, while for sink absorption the RC time displays a non-trivial scaling $\propto k^{-1/2}$, due to the fact that most reaction events occur in a small penetration length from the target surface.

These results for both models (i) and (ii) have been confirmed by numerical simulations for confining volumes of various shapes (see Fig. \ref{FigDiffusion}), which have been  chosen as representative of non-spherical volumes, displaying anisotropy ($A$) or protrusions ($B$). Importantly,  numerical results confirm our   prediction  that for $d\ge 2$ the full distribution is still given by  Eq. (\ref{SurvNonCompact})  for both continuous models (where the case $d=2$ is considered as non compact) [Fig \ref{FigDiffusion}(f-g)].  

\section*{Conclusions} 

We have provided a  general formalism to determine the  reaction time distribution for imperfect reactions involving  the broad class of diffusive and anomalously diffusive Markovian transport processes  in confinement. We have investigated several representative mechanisms of imperfect reactivity to test the robustness of our conclusions. Importantly, our results show that  the first moment alone, although not representative of typical reaction times,  gives access to the full distribution in the large volume limit, which allows to quantify reaction kinetics at all timescales. 
Thanks to this property, our formalism can be adapted in principle to refined  mechanisms of imperfect reactivity (gating, orientational constraints...), as soon as  the mean first passage time can be asymptotically determined. Remarkably, and counter-intuitively, we find that in the large volume limit the reaction time distribution is  identical to that of the first-passage time upon an appropriate rescaling of parameters. This implies that for compact transport processes, the reaction time distribution is   broadly distributed with large fluctuations even  in the reaction controlled regime where the mean reaction time is independent of the transport process. This is in striking contrast with the  naive prediction of exponentially  distributed reaction times  for  first-order kinetics, which in fact is valid only for extremely low reactivity. This unexpected property could lead to large  fluctuations of concentrations -- as observed in the context of gene expression -- even in simple reaction schemes, and even for  low  reactivity. We expect that the main effect identified here, i.e. that complex first passage properties due to compact transport do not disappear for imperfect reactivity, could be could be extended  to more general processes that are more complex than scale invariant Markovian processes, and to more complex reactions schemes potentially involving competitive reactions. This will be the subject of future works.
 
 \vspace{0.5cm}
{\bf Data availability.}   
  The numerical data presented in Figures 2 and 3 are available from the corresponding author on reasonable request.

   \vspace{0.5cm}
   {\bf Code availability.}   
 The code that generated the data presented in Figures 2 and 3 is available from the corresponding author on reasonable request.

     \vspace{0.5cm}
{\bf Author contributions.}   
  All authors (T. G., M. D., O. B., R. V.) contributed equally to this work.
  
   \vspace{0.5cm}
{\bf Competing interests.}   
  The authors declare no competing interests.
 
 \begin{acknowledgments}
Computer time for this study was provided by the computing facilities MCIA (Mesocentre de Calcul Intensif Aquitain) of the Universit\'e de Bordeaux and of the Universit\'e de Pau et des Pays de l'Adour. We thank J\'er\'emie Klinger for providing his codes for the simulations of random walks on critical percolation clusters. 
\end{acknowledgments}

\newpage

\renewcommand{\thetable}{  \textbf{Supplementary Table S\arabic{table}}}
\renewcommand{\tablename}{}
\renewcommand{\thefigure}{ \textbf{Supplementary Figure S\arabic{figure}}}
\renewcommand{\figurename}{}
 \renewcommand{\theequation}{S\arabic{equation}}
 \renewcommand{\refname}{Supplementary References}
 \def\bibsection{\section*{\refname}} 

\setcounter{equation}{0} 
\setcounter{figure}{0} 
\renewcommand*{\citenumfont}[1]{S#1}
\renewcommand*{\bibnumfmt}[1]{[S#1]}

\onecolumngrid
\appendix
%
\section*{Supplementary Information}
%
%
%
%
%
%

In this Supplementary Information, we provide:
\begin{itemize}
\item A brief derivation of Eqs. (2) and (3) in the main text (Supplementary Note 1).
\item Details on the derivation of Eq. (7) in the main text, for the calculation of the distribution of reaction times for compact searches (Supplementary Note 2).
\item Details on reaction times on fractal networks  (Supplementary Note 3), including the method we used to calculate the distribution of first reaction times on large deterministic fractal networks, additional results on Vicsek fractals (Supplementary Figure~S1),  additional results for other fractals (Supplementary Figure~S2), details on simulations on the percolation cluster and  a summary table of fractal dimensions for all networks considered in this work (Supplementary Table~S1). 
\end{itemize}

\subsection*{Supplementary Note 1: Derivation of Eqs. (2),(3) in the main text}
\label{Section90432}
We start from Eq.(1) in the main text:
\begin{align}\label{F_de_t_SI}
 F(T  & \vert \ve[r]  ,p )=
  \sum_{n=1}^{\infty}  \int_0^\infty\mathrm{d}\tau_{\ve[r]} \left[\prod_{k=2}^n \int_0^\infty d\tau_1^{(k)} F_1^*(\tau_1^{(k)})  \right] 
p(1-p)^{n-1}  F^*(\tau_{\ve[r]}  | \ve[r]) \delta\left(T-\tau_{\ve[r]}-\sum_{k=2}^n   \tau_1^{(k)}  \right),
\end{align}
Now, taking the Laplace transform with respect to the variable $T$, we obtain
\begin{align}
 \tilde{F}(s  & \vert \ve[r]  ,p )=
\int_0^\infty dT e^{-sT}  \sum_{n=1}^{\infty}  \int_0^\infty\mathrm{d}\tau_{\ve[r]} \left[\prod_{k=2}^n \int_0^\infty d\tau_1^{(k)} F_1^*(\tau_1^{(k)})  \right] p(1-p)^{n-1}  F^*(\tau_{\ve[r]}  | \ve[r]) \delta\left(T-\tau_{\ve[r]}-\sum_{k=2}^n   \tau_1^{(k)}  \right),
\end{align}
We change the order of integration and integrate with respect to $T$ first:
\begin{align}
 \tilde{F}(s  & \vert \ve[r]  ,p )=
\sum_{n=1}^{\infty}  \int_0^\infty\mathrm{d}\tau_{\ve[r]} \left[\prod_{k=2}^n \int_0^\infty d\tau_1^{(k)} F_1^*(\tau_1^{(k)})  \right] p(1-p)^{n-1}  F^*(\tau_{\ve[r]}  | \ve[r]) e^{-s\tau_{\ve[r]}-s\sum_{k=2}^n   \tau_1^{(k)} }  
\end{align}
The integrals factorize, leading to
\begin{align}
 \tilde{F}(s  & \vert \ve[r]  ,p )=
\sum_{n=1}^{\infty}   [\tilde{F}_1^*(s)]^{n-1} p(1-p)^{n-1}  \tilde{F}^*(s  | \ve[r])  \label{03424}
\end{align}
We recognize a geometrical series and we thus obtain
\begin{align}
 \tilde{F}(s  & \vert \ve[r]  ,p )=  \frac{p \tilde{F}^*(s  | \ve[r])}{1-\tilde{F}_1^*(s)(1-p)}
\end{align}
which is Eq. (2). Now, using the small-$s$ expansions
\begin{align}
\tilde{F}^*(s  | \ve[r])=1-s \langle \tau_{\ve[r]}\rangle+\mathcal{O}(s^2), \hspace{0.5cm}
\tilde{F}_1^*(s  )=1-s \langle \tau_1\rangle+\mathcal{O}(s^2),  \hspace{0.5cm}
\tilde{F}(s    \vert \ve[r]  ,p )=1-s \langle T_{\ve[r]}(p)\rangle+\mathcal{O}(s^2),
\end{align}
we see that an expansion of Supplementary Equation (\ref{03424}) at linear order in $s$ leads to
\begin{equation}
\langle T_{\ve[r]}(p)\rangle=\langle\tau_{\ve[r]} \rangle+\frac{1-p}{p} \langle \tau_1\rangle  \label{meanRT_SI} .
\end{equation}
which is Eq. (3). 

\subsection*{Supplementary Note 2: Derivation of Eq. (7): Distribution of first reaction times for compact searches}
\label{DistriFPT}
Here we identify the distribution of reaction times in discrete fractal networks for compact searches. Our strategy is to identify all its moments by exploiting the fact that $\tilde{F}(s\vert r,p)$ is the generating function of the moments, i.e.  
\begin{align}
\tilde{F}(s\vert r,p)=\langle e^{-sT_r(p)}\rangle=\sum_{n=0}^\infty (-1)^n\   s^n \langle T_r^n(p)\rangle /n!  
\end{align}
Using the fact that a similar relation can be written for all distributions, we write Eq.~(3) (in the main text) by using the expansion of  $\tilde{F}^*$ and $\tilde{F}_1^*$ as an infinite series (involving the moments $\langle\tau_r^n\rangle$ and $\langle\tau_1^n\rangle$) and use the expansion of the function $1/(1-x)$ near $x=0$ to obtain
 \begin{align}
\tilde{F}(s|r,p)=\frac{p\ \tilde{F}^{*}(s|r) }{1-(1-p)\tilde{F}_1^{*}(s) }=\left[1+\sum_{q=1}^\infty \frac{(-1)^q \ s^q \langle \tau_r^q\rangle}{q!}\right]\times  
\left\{1+ \sum_{m=1}^\infty    \left[\sum_{n=1}^\infty\frac{(1-p)(-1)^n}{p \ n!} \langle \tau_1^n \rangle s^n  \right]^m\right\} \label{Series49231}.
\end{align}
At this stage, identifying all moments $\langle T_r^n(p)\rangle$ (thus the coefficient of $s^n$ for any $n$ in the above expression) seems an intractable task. However, we recall that the moments of the mean first passage time are known \cite{Benichou2010_SI,levernier2018universal} to scale as 
\begin{align}
\langle \tau_r^n\rangle \sim a_n r^{d_w-d_f}R^{d_f+(n-1) d_w} \label{ScalingMomentsFPT},
\end{align}
where the coefficients $a_n$ do not depend on the geometry. Similar scaling (with $r=1$), holds for $\langle \tau_1^n\rangle$. These scalings can be used to show that products of moments are negligible compared to moments of higher order, for example the following estimate
\begin{align}
\frac{1}{p^M}\langle \tau_1^{n_1}\rangle\langle \tau_1^{n_2}\rangle...\langle\tau_1^{n_M} \rangle   \ll\frac{1}{p }  \langle \tau_1^{n_1+n_2+...+n_M}\rangle \label{94321}
\end{align}
holds when $R\gg l_c$ (for any $M\ge2$ for which all $n_i\ge1$); and the following  relation is also valid:
\begin{align}
 \langle \tau_r^{n_1}\rangle\langle \tau_1^{n_2}\rangle \underset{r\ll R}{\ll} \langle \tau_r^{n_1+n_2}\rangle \label{94314}.
\end{align}
Using Supplementary Equation (\ref{94321}), we see that for all $n$, the terms coming from the $m=1$ term of the series dominate all the others (because all coefficients of $s^n$ generated by products of moments are negligible compared to the corresponding term coming from $m=1$ in the series). This leads to 
 \begin{align}
\tilde{F}(s|r,p)\simeq \left[1+ \sum_{q=1}^\infty \frac{(-1)^q \ s^q \langle \tau_r^q\rangle}{q!}\right]\times  
\left\{1+    \sum_{n=1}^\infty\frac{(1-p)(-1)^n}{p \ n!} \langle \tau_1^n \rangle s^n \right\} .
\end{align}
Next, using Supplementary Equation (\ref{94314}) we see that all products of moments are negligible compared to terms that are not products of moments, so that 
 \begin{align}
\tilde{F}(s|r,p) \simeq 1+ \sum_{q=1}^\infty \frac{(-1)^q \langle \tau_r^q\rangle}{q!} s^q+  
    \sum_{n=1}^\infty\frac{(1-p)(-1)^n}{p \ n!} \langle \tau_1^n \rangle s^n . 
\end{align}
This means that for all $q\ge1$
\begin{align}
\langle T_r^n(p)\rangle\underset{R\to\infty}{\sim}  \langle  \tau_r^n\rangle+\frac{  (1-p) }{p }\langle \tau_1^n \rangle . 
\end{align}
Up to now, the only approximation is the large volume limit. Let us do another approximation: we assume that that the scaling Supplementary Equation (\ref{ScalingMomentsFPT}) holds also for $r=1$ ; such approximation has proved accurate for the first moment \cite{benichou2008zero_SI}. In this case we get
\begin{align}
\langle T_r^n(p) \rangle\simeq  a_n \left[r^{d_w-d_f} +\frac{  (1-p) }{p }\right] R^{d_f}R^{d_w(n-1)} . 
\end{align}
We realize that the above relation can also be written as 
\begin{align}
\langle T_r^n(p)\rangle \simeq \frac{\langle T_r(p)\rangle}{\langle\tau_r\rangle}\langle\tau_r^n\rangle\label{9313},
\end{align}
which means that all moments of the reaction time are proportional to the moments of the first passage time with a proportionality factor which does not depend on $n$. This implies a proportionality between the two distributions: 
\begin{align}
F( t \vert r,p ) \simeq \frac{\langle T_r(p) \rangle}{\langle\tau_r\rangle} F^*(t\vert r) . 
\end{align}
Using the FPT distribution given in Ref. \cite{Benichou2010_SI} finally leads to Eq. (7) in the main text. Note finally that Supplementary Equation (\ref{9313}) holds in both limits of strong and weak reactivity, so that the hypothesis Supplementary Equation (\ref{ScalingMomentsFPT}) is required only in the crossover regime. 
 
\subsection*{Supplementary Note 3: Additional details on the distribution of reaction times on fractal networks}
\label{DetailFractal}
\subsubsection{Details on the method to obtain reaction time distribution on large deterministic fractal networks}\label{SEctionMethod}
We consider the dynamics of a random walker on a network of $N$ sites and connectivity matrix $M$, such that $M_{ij}=-1$ when sites $i$ and $j$ are connected and the diagonal elements are $M_{ii}=f_i$, with $f_i$ the functionality, or connectivity of site $i$ (i.e. the number of linked neighbors). Let us consider a random walker on this network, and let us call $p_i(t)$ be the probability to find it at time $t$ at site $i$. We assume that, between $t$ and $t+dt$, there is a probability $\mu \ dt $ (for each edge) to jump on this edge, so that the average waiting time on site $i$ is $(f_i\mu)^{-1}$). If we consider also one reactive site $i_0$, which is absorbing with rate $k$ (imperfect reactivity), the master equation for this dynamics  is 
\begin{align}
\partial_t \ve[p] =-M\ve[p]-k\  \ve[u] \ (^t\ve[u]\cdot \ve[p]) . 
\end{align}
where $\ve[u]$ is a column vector which encodes the position of the reactive site ($u_i=0$ for all $i$ except for $u_{i_0}=1$), and ${\ve[p]=\ ^t(p_1,p_2...)}$. Note that we have chosen the units of time so that $\mu=1$. With this dynamics, the stationary probability in absence of reaction is $p_i=1/N$ so that there is no confusion between averages over stationary configurations and uniform averages. $M$ is symmetric so that it can be expressed as $M=Q D Q^{-1}$ with $Q$ an orthonormal matrix ($Q^{-1}=\ ^tQ$) and $D$ a diagonal matrix of positive eigenvalues $0=\lambda_1<\lambda_2<...<\lambda_N$ (possibly degenerate).  We pose 
\begin{align}
\ve[q]=Q^{-1}\ve[p], \hspace{2cm}\ve[v]=Q^{-1}\ve[u].
\end{align}
The dynamics in the space of eigenmodes reads 
\begin{align}
\partial_t \ve[q] =-D\ve[q]-k \ve[v] \ (^t\ve[v]\cdot \ve[q]) \label{Eq_q}. 
\end{align}
Now, the distribution of first reaction times is simply $F(t)=k \ (^t\ve[u]\cdot \ve[p]) =k \ (^t\ve[v]\cdot \ve[q])$. Taking the Laplace transform of Supplementary Equation (\ref{Eq_q}) leads (after a few manipulations) to  
\begin{align}
\tilde{F}(s) = \frac{1}{1+\frac{k}{sN} +k \sum_{  \lambda\neq0}  \frac{1}{s+\lambda} \sum_{i\in   I(\lambda)} v_i^2 }  \left[   \frac{k}{sN}+k \sum_{ \text{distinct } \lambda\neq0}  \frac{1}{s+\lambda} \sum_{i\in   I(\lambda)}  v_i q_i(0)  \right],
\end{align}
where we have decomposed the sum over distinct values of $\lambda$ different from zero, and $I(\lambda)$ represents the ensemble of indexes $i$ so that $\lambda_i=\lambda$. Now, the key point is that $ \sum_{i\in   I(\lambda)} v_i^2 $ is actually the squared norm of the projection $P_\lambda  \ve[u]  $  over the eigensubspace associated to $\lambda$. Similarly, $\sum_{i\in   I(\lambda)}   v_i q_i(0) $ is the scalar product between $P_\lambda \ve[u]$ and the projection $P_\lambda   \ve[p](t=0)$ of the vector of initial probabilities over the eigensubspace associated to $\lambda$. 
Although brute force diagonalization of the connectivity matrix $M$ is in practice limited to a few thousands sites $N$, we can implement a procedure to identify iteratively (from generation $g$ to $g+1$) all the eigensubspaces. We refer to Ref. \cite{dolgushev2015contact} for details of such iterative procedures for (i) the case of Vicsek fractals and (ii) the dual Sierpinski gasket. In practice, these projections can be computed relatively cheaply (within a few days on a single processor) up to $g=13$ (dual Sierpinski gasket, so that $N\simeq 1.6 \times 10^6 $ sites), and for Vicsek fractals $g=8,7,6$ for global functionalities $f=3,4,6$, respectively.   

Note that calculating these projections for a given vector $\ve[u]$ and $\ve[p](t=0)$ thus gives access to the whole probability distribution in Laplace space, thus for all times after numerical Laplace inversion, for all values of the reactivity parameter $k$. In particular the first moments can be expressed as 
\begin{align}
\langle T\rangle=\frac{N}{k} + \sum_{\text{distinct }\lambda\neq0}\frac{N}{\lambda}\sum_{i\in I(\lambda)} [v_i^2     -   v_i q_i(0)].
\end{align}
Note that the global mean reaction time is available by choosing the stationary distribution for $ p_i(t=0)=1/N$. Finally we note that the  probability $p$ to be absorbed at each visit of the target in this model is 
\begin{align}
p=\int_0^\infty dt \ k \ e^{-(k+f_{i_0})t}=\frac{k}{k+f_{i_0}}. 
\end{align}

\subsubsection{Results for Vicsek fractals of different functionalities}

\begin{figure}[h]
 \centering
\includegraphics[width= \linewidth]{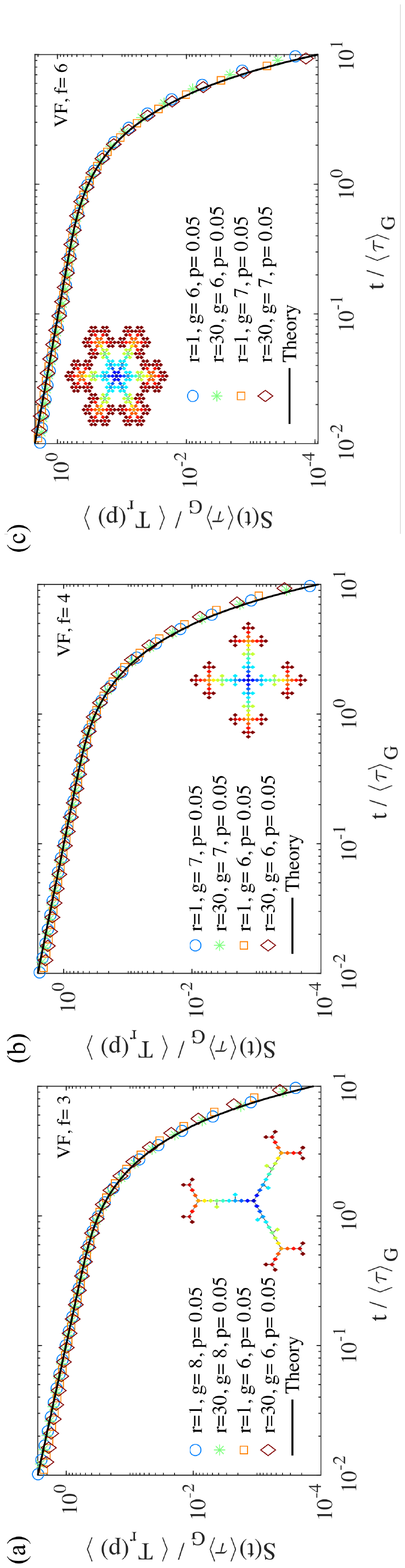}
 \caption{Additional results obtained with the method described in Supplementary Note 3.a for Vicsek fractals of different functionalities, (a): $f=3$, (b): $f=4$; (c) $f=6$. In each graph, the network is represented in inset at generation $g=3$.  
 }\label{FigVF}
\end{figure}

For Vicsek fractals, Different values of $d_w$ and $d_f$ can be explored by varying the functionality of the central bead. In the main text, we show the results for $f=6$, additional examples are shown on\ref{FigVF} and show that our theory is confirmed for different values of $f$. 

\subsubsection{Results of stochastic simulations for other fractal networks}

\label{SimusFractal}

In the main text, we have considered a dynamics on networks  for which the average waiting time on each site is inversely proportional to its number of neighbors. Here we consider ``traditional'' random walk simulations, in which the waiting time at each site is uniform and taken as unity. We have performed stochastic simulations for this dynamics and we have checked that this different type of dynamics does not change the validity of our results.  The figure below demonstrates that, once properly rescaled by appropriate first moments, the shape of the survival probability falls into the universality classes predicted in the main text. 
 
\begin{figure}[h]
\vspace{-0.3cm}
\centering
\includegraphics[width=0.32\linewidth,trim=0 0 0 0]{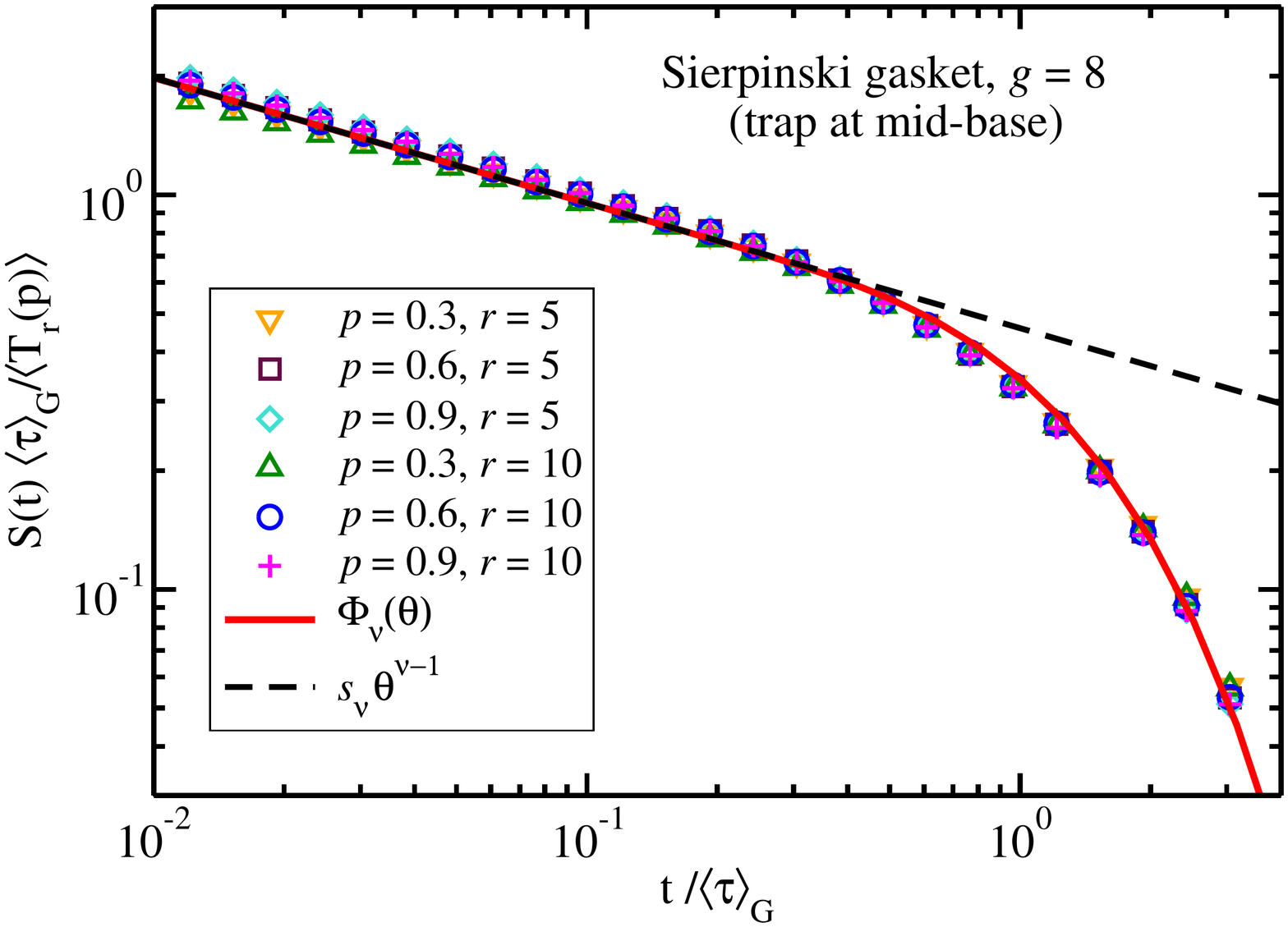}
\includegraphics[width=0.32\linewidth,trim=0 0 0 0]{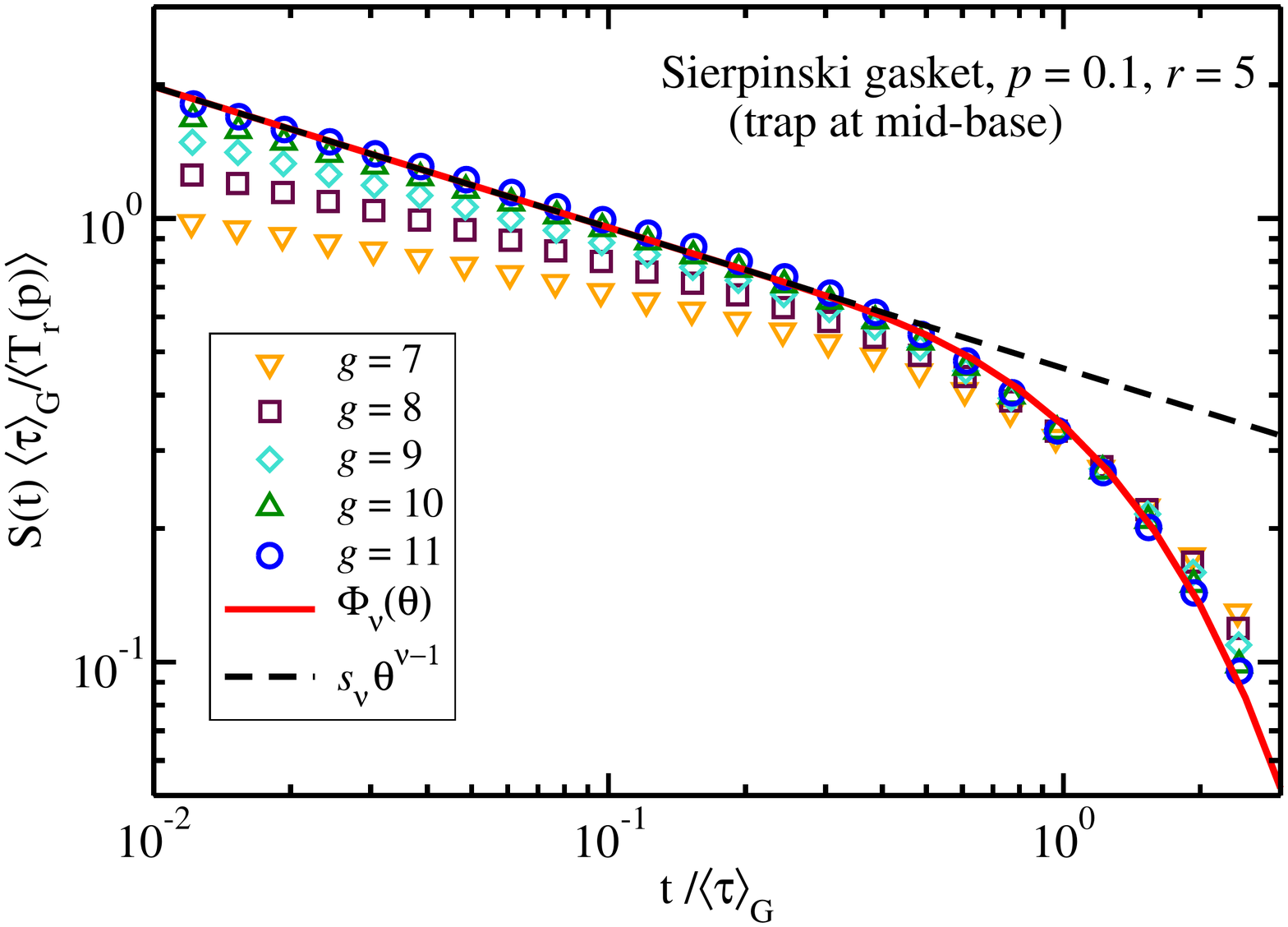}
\includegraphics[width=0.32\linewidth,trim=0 0 0 0]{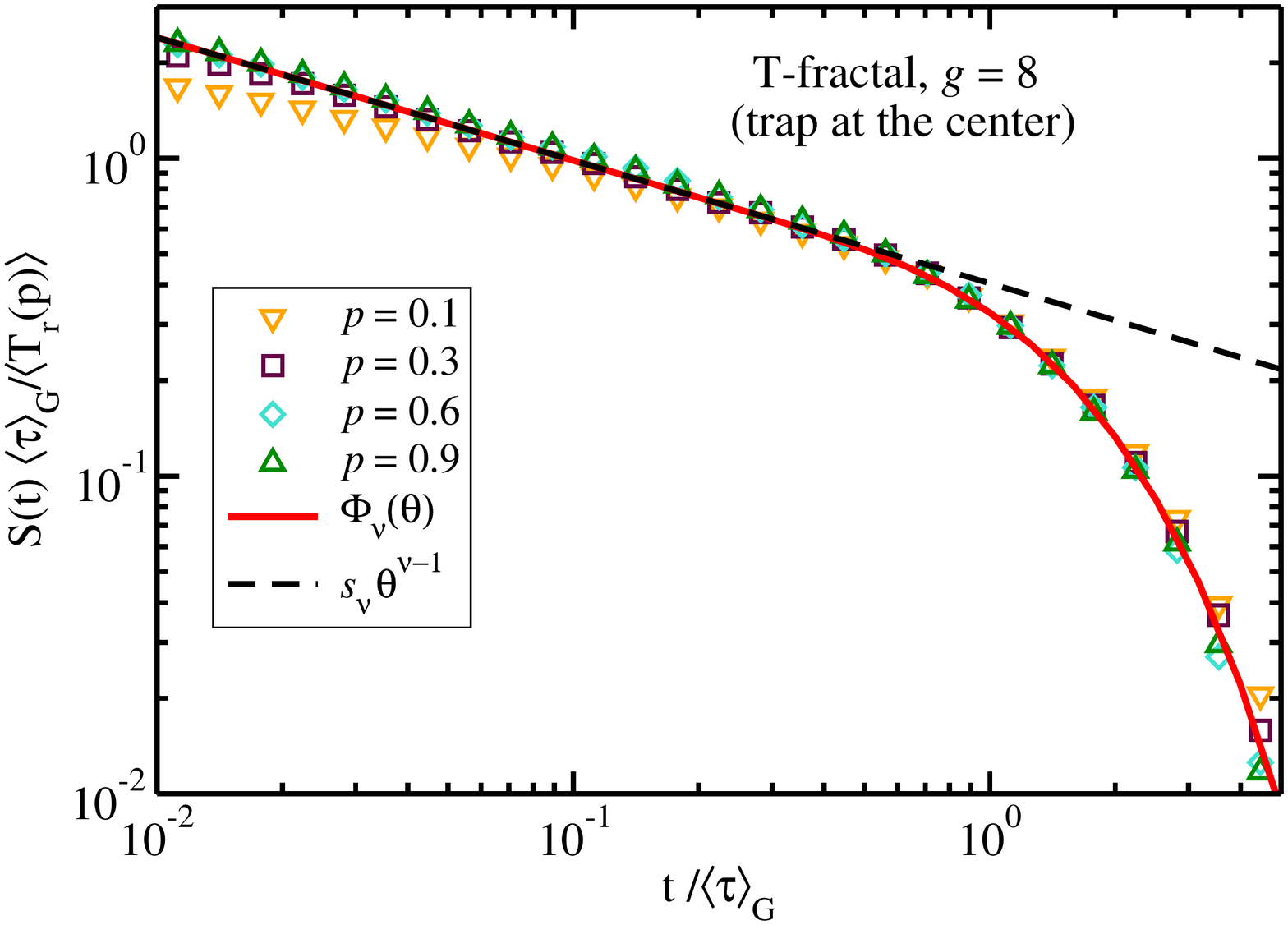}
\caption{Rescaled survival probability for random walks on Sierpinski gasket (for which $\nu=d_f/d_w=\ln 3/\ln 5$) and on T-fractal ($\nu=\ln 3/\ln 6$). The red line represents $\Phi_\nu(\theta)$ given in Eq. (8) of the main text, whose algebraic decay (dashed line) contains the prefactor $s_{\nu}= \left(\frac{2\nu}{1-\nu^2}\right)^{\nu}\frac{\nu}{\Gamma(1-\nu)}$. For both fractals the global mean first passage time $\langle\tau\rangle_{\mathrm G}$ used for rescaling of time [see Eq. (7) of the main text] is known analytically \cite{Agliari2008,Haynes2008}; for T-fractal the simulation data is averaged over chemical distances $r\in[1,32]$. For the graph in the middle, $p=0.1$ is fixed, note the convergence towards the predicted value when the size of the system increases ($g$ is the generation number).  
\vspace{-0.3cm}
}\label{FigSG}
\end{figure} 
  
\subsubsection{Details on simulations for random walks on the percolation cluster}
To generate a percolation cluster, we have used a  regular $200\times200$ periodic two-dimensional square lattice on which we have removed randomly half of the bonds. Then, we have identified the connected network of maximal size (with the algorithm of Ref. \cite{newman2001fast}) on which random walks simulations were performed, by prescribing that the waiting time at each site is inversely proportional to its connectivity, so that the uniform distribution is also the equilibrium distribution. For each run, the target   and the initial position were chosen uniformly with the constraint of fixed (chemical) distance between them. To generate Fig. 2(e) of the main text, the MRT was estimated from Eq. (5), and the GMFPT was identified numerically for the percolation cluster under consideration. 

  
\begin{table}
  \caption{Summary of fractal dimensions for all fractal networks considered in this work}
\begin{center} 
\begin{tabular}{|c|r|r|}
  \hline
  Fractal & Spatial dimension ($d_f$)   & Walk dimension ($d_w$) \\
  \hline
  Vicsek fractal (f=3) 			&	 $\ln(1+3)/\ln3\simeq 1.26 $ & $1+d_f\simeq 2.26 $ \\
    Vicsek fractal (f=4) 		& $\ln(1+4)/\ln3\simeq 1.47 $ 	& $1+d_f\simeq 2.47 $ \\
      Vicsek fractal (f=6) 		& $\ln(1+6)/\ln3\simeq 1.77 $ 	& $1+d_f\simeq 2.77 $ \\
  Sierpinski Gasket 			 & $\ln3/\ln2\simeq 1.59$ & $\ln5/\ln2\simeq 2.32$ \\
Dual Sierpinski Gasket (DSG) 	& $\ln3/\ln2\simeq 1.59$ 			& $\ln5/\ln2\simeq 2.32$ \\  
T-fractal 					& $ \ln 3/\ln 2\simeq  1.59 $	& $ \ln 6 /\ln 2 \simeq 2.59 $ \\
2D (bond) Percolation cluster 	& $ 91/48\simeq 1.90 $  		& 2.878 \\
  \hline
\end{tabular}
  \end{center}
   \end{table}
 



\end{document}